# Phase response reconstruction for non-minimum phase systems using frequency-domain magnitude values

Taco Boland[1*], Rik Naus[2], Peter Zwamborn[1]

[1]TNO Electronic Defence, Oude Waalsdorperweg 63, 2597 AK, The Hague, The Netherlands
[2]TNO Quantum Technology, Stieltjesweg 1, 2628 CK, Delft, The Netherlands
*taco.boland@tno.nl

**Abstract:** It is often more complicated to measure the phase response of a large system than the magnitude. In that case, one can attempt to use the Kramers-Kronig (KK) relations for magnitude and phase, which relates magnitude and phase analytically. The advantage is that then only the magnitude of the frequency response needs to be measured. We show that the KK relations for magnitude and phase may yield invalid results when the transfer function has zeros located in the right half of the complex $s$-plane, i.e. the system is non-minimum phase. In this paper we propose a method which enables to determine these zeros, by using specific excitation signals and measuring the resulting time responses of the system. The method is verified using blind tests among the authors. When the locations of the zeros in the right half of the complex $s$-plane are known, modified KK relations can be successfully applied to non-minimum phase systems. We demonstrate this by computing the phase response of the electric field, excited by a point dipole source inside a closed cavity with Perfect Electrically Conducting (PEC) walls. Also, the effects of simulated measurement noise are considered for this example.

## 1. Introduction

The frequency response of a system is an important property to define the relation between the input and output and is used for characterizing and designing systems. The frequency response generally consists of a magnitude and a phase. This is under the assumption that the system behaves (approximately) Linear Time Invariant (LTI).

Applications where measuring the frequency response is important are among others in the Defence industry. An example is the evaluation of navy ships for hardening against the Nuclear Electro Magnetic Pulse (NEMP) threat. The frequency response of the electric field transmission, from a RF source outside, is measured at key positions inside the ship. Using the measured frequency response, a full-threat (N)EMP can then be numerically simulated in the time domain by the Inverse Fast Fourier Transform (IFFT). This has considerable advantages in terms of costs, complexity and safety measures compared to generating a full-threat EMP and measuring the time response of the electric field inside the ship. A problem which emerges in this application, is that the phase of the frequency response cannot be measured easily. However, the phase information is also required to accurately reconstruct the time response of the electric field inside the ship when a (N)EMP threat is simulated. Other applications, where the phase is important but difficult to measure, can be found in the areas such as optical device characterization and biomedical imaging [1].

Fortunately, the magnitude of the frequency response can often be measured accurately. To avoid the necessity to use the phase information, one could in some cases use a minimum phase method as described in e.g. [2]. In [3] [4] [5] the KK relations for magnitude and phase are used to accurately calculate the phase from the magnitude of the frequency response. This is under the assumption that the corresponding system is minimum phase. It is equivalent to requiring that the transfer function of the system does not have zeros in the right half of the complex $s$-plane. However, the minimum phase method is not always applicable and some examples of non-minimum phase systems are described in [6]. One of these examples is the optical reflection coefficient of optical materials. Another example describes the frequency response of mechanical systems such as an atomic force microscope which use a flexible cantilever to probe a surface.

In [3] [5] modifications of the KK relations, using the Blaschke product, are described. These modifications allow the use of the KK relations for magnitude and phase, even when the system is non-minimum phase. However, an important assumption here is that the locations of the zeros of the transfer function in the right half of the $s$-plane are known. It is stressed that for many applications the locations of these zeros are unknown, and difficult to obtain.

Here we propose a method which enables us to accurately estimate the locations of the zeros in the right half of the $s$-plane. This is done using specific excitation signals and calculating the time responses of the system. For this method, no a priori knowledge of the transfer function is required. After using our new zero-search method to precisely locate the zeros, we then successfully reconstruct the phase from the measured magnitude of the frequency response using modified KK relations.

This paper is organized as follows: Section 2 gives a detailed mathematical description of the EM cavity model, where we apply the theory described in [7] to a closed PEC cavity with a point dipole source. Section 3 discusses the KK relations for magnitude and phase and is concluded with some numerical results where the KK relations for magnitude and phase are applied to the cavity model from Section 2. In Section 4 modified KK relations using the Blaschke product are introduced. In Section 5 we describe our new zero-search method including numerical results. It is concluded with a successful application of modified KK relations to the cavity model from Section 2. In Section 6 we investigate the consequences of noise on the accuracy of the proposed zero-search method and the phase reconstruction method using the modified KK relations. An example with added noise using the cavity model from Section 2 is discussed. Finally, we present conclusions and recommendations.

## 2. The cavity model

### 2.1. Geometry

We start with defining the geometry of the problem under consideration, which we will use to test the phase reconstruction using the KK relations for magnitude and phase. The geometry consists of a closed cavity with PEC walls and enclosed current source $J$. An example of such a cavity is a rectangular cavity with dimensions $a$, $b$ and $d$, is shown in *Fig. 1*.

### 2.2. Eigenmode expansion of the electric field

The boundary conditions of the electric field on the perfectly conducting cavity walls, $\partial\Omega_{wall}$, are given by

$$\mathbf{n} \times \mathbf{E}(\mathbf{r}, t) = 0, \qquad (1)$$

where $\mathbf{n}$ is the inward normal vector on $\partial\Omega_{wall}$.

The charge free domain $\Omega$ inside the cavity is filled with a homogeneous isotropic time-invariant insulator. Thus, the electric field $\mathbf{E}(\mathbf{r}, t)$ at location $\mathbf{r}$ inside the cavity and at time $t$, satisfies the equation

$$\nabla^2 \mathbf{E}(\mathbf{r}, t) = \epsilon\mu \frac{\partial^2}{\partial t^2} \mathbf{E}(\mathbf{r}, t) + \mu \frac{\partial}{\partial t} \mathbf{J}(\mathbf{r}, t). \qquad (2)$$

In this equation, $\nabla^2$ denotes the vector Laplace operator, $\epsilon$ and $\mu$ denote the permittivity and permeability of the insulator, respectively and $\mathbf{J}(\mathbf{r}, t)$ denotes the impressed electric current density inside $\Omega$.

Since we consider the field distribution inside the cavity, it is advantageous to describe the electric field, $\mathbf{E}(\mathbf{r}, t)$, in terms of its truncated modal expansion [7]. For a rectangular cavity, it is convenient to make a distinction between transverse magnetic $TM$ and transverse electric $TE$ modes. This means that either the electric or the magnetic field is forced to be transverse to the $z$ axis in our Cartesian coordinate system. The modal expansion is split up in a transverse magnetic and a transverse electric part

$$\mathbf{E}(\mathbf{r}, t) = \mathbf{E}_{TM}(\mathbf{r}, t) + \mathbf{E}_{TE}(\mathbf{r}, t) \qquad (3)$$

$$\approx \sum_{v=1}^{N_{TM}} \mathbf{E}_v(\mathbf{r}) x_v(t) + \sum_{v=N_{TM}+1}^{N_{TM}+N_{TE}} \mathbf{E}_v(\mathbf{r}) x_v(t),$$

where $N_{TM}$ denotes the number of TM modes $N_{TE}$ denotes the number of TE modes used in the modal expansion, $\mathbf{E}_v(\mathbf{r})$ is the 3-D electric field distribution of the $v$-th mode of the cavity and $x_v(t)$ is the corresponding time dependent expansion coefficient. The total number of modes used is $N = N_{TM} + N_{TE}$. For an exact representation, an infinite number of modes should be considered: $N \to \infty$.

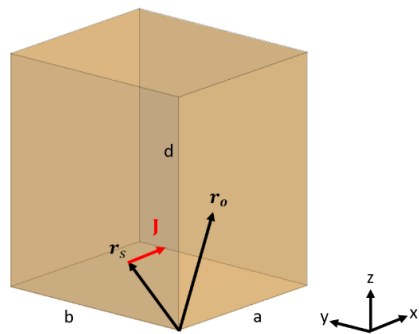

***Fig. 1.*** *Rectangular cavity geometry with dimension a, b and d. The current source $\mathbf{J}(\mathbf{r}, s)$ is located at $\mathbf{r} = \mathbf{r}_s$ and the observation point is located at $\mathbf{r} = \mathbf{r}_o$.*

However, for numerical purposes the number of modes taken should be finite but sufficiently large so that the electric field is computed appropriately.

The 3-D electric field distributions of the modes $\mathbf{E}_v(\mathbf{r})$ can be found by solving the source-free 3-D Helmholtz equation in the frequency domain and enforcing the boundary conditions in (1)

$$\nabla^2 \mathbf{E}_v(\mathbf{r}) + k_v^2 \mathbf{E}_v(\mathbf{r}) = 0, \qquad (4)$$

where $k_v^2 = \omega_v^2 \epsilon \mu$ is the $v$-th wavenumber corresponding to the $v$-th mode which will resonate with frequency $\omega_v$.

Here, the $TM$ and $TE$ mode field patterns of a rectangular cavity are calculated analytically [8]. Note, however, that the method described in this paper can also be applied to a closed cavity with any geometry, if the modes of the cavity can be found by solving (4) analytically or numerically.

### 2.3. Cavity transfer function

The cavity transfer function defines the frequency response of the electric field at an arbitrary location due to a local current source inside the cavity. To obtain the cavity transfer function we consider a source located at $\mathbf{r} = \mathbf{r}_s$ and an observation point located at $\mathbf{r} = \mathbf{r}_o$ inside the cavity and find an explicit relation between the impressed current density $\mathbf{J}(\mathbf{r}, t)$ and the observed field $\mathbf{E}(\mathbf{r}, t)$.

To keep the model simple, we assume that $\mathbf{J}(\mathbf{r}, t)$ in (2) is a single dipole source, oriented in the $x$-direction as shown in *Fig. 1*. Furthermore, we assume that the dipole has length $L$ [m], and has a uniform current distribution. Thus, we define

$$\mathbf{J}(\mathbf{r}, \mathbf{r}_s, t) = -\hat{\mathbf{x}} f(\mathbf{r} - \mathbf{r}_s) i(t) \qquad (5)$$

$$f(\mathbf{r} - \mathbf{r}_s) = L \Pi_L(x - x_s) \delta(y - y_s) \delta(z - z_s)$$

$$\Pi_L(x - x_s) = \begin{cases} 0 & |x - x_s| > \frac{L}{2} \\ \frac{1}{L} & |x - x_s| \leq \frac{L}{2}, \end{cases}$$

where $\hat{\mathbf{x}}$ is a unit vector in the $x$-direction, $i(t)$ [A] the impressed current source function, which is the input current of the dipole and $\delta(x)$ is the Dirac delta function. The function $f(r - r_s)$ is defined such that the current distribution is localised within an infinitesimal volume hence $L \downarrow 0$.

In [7], it is discussed to rewrite (2) as a system of second order Ordinary Differential Equations (ODEs) by substituting (5) in (2) and using the orthogonality of the cavity mode field patterns,

$$2W_\nu \left( \omega_\nu^2 x_\nu(t) + \frac{\partial^2 x_n(t)}{\partial t^2} \right) = -\frac{\partial}{\partial t} \iiint_\Omega E_\nu(r) \cdot J(r,t) dr, \quad (6)$$

where $W_\nu$ indicates the energy stored in the $\nu$-th mode, which is computed as

$$W_\nu = \frac{\epsilon}{2} \iiint_\Omega E_\nu(r) \cdot E_\nu^*(r) dr \quad (7)$$

where $\epsilon$ is the permittivity of the medium inside the cavity and $*$ denotes complex conjugation.

While assuming that $E_\nu(r)$ is constant across the infinitesimal volume enclosing $J(t, r, r_s)$, we substitute (5) into (6) which leads to

$$2W_\nu \left( \omega_\nu^2 x_\nu(t) + \frac{\partial^2 x_\nu(t)}{\partial t^2} \right) = L[E_\nu(r_s)]_x \frac{\partial}{\partial t} i(t). \quad (8)$$

Equation (8) is integrated with respect to time, yielding

$$\omega_\nu^2 \dot{x}_\nu(t) + \frac{\partial x_\nu(t)}{\partial t} = \frac{L}{2W_\nu}[E_\nu(r_s)]_x i(t), \quad (9)$$

$$\frac{\partial \dot{x}_\nu(t)}{\partial t} = x_\nu(t),$$

where $[E_\nu(r_s)]_x$ denotes the $x$ component of the $\nu$-th mode field pattern at the location of the source.

After some mathematical steps, a commonly known state equation is obtained from (9)

$$\frac{\partial}{\partial t} x(t) = A x(t) + b i(t). \quad (10)$$

with input $i(t)$, input vector $b$, state matrix $A$ and state vector $x(t)$ defined as

$$x(t) = [\dot{x}_1(t) \quad x_1(t) \quad \cdots \quad \dot{x}_N(t) \quad x_N(t)]. \quad (11)$$

Solving this equation for $x(t)$ leads to the time dependent expansion coefficients in (3). Now $E(r_o, t)$, the electric field inside the cavity observed at position $r_o$, can be written in terms of the state vector $x(t)$

$$E(r_o, t) = [0 \quad E_1(r_o) \quad \cdots \quad 0 \quad E_N(r_o)] x(t), \quad (12)$$

which we rewrite as $E(r_o, t) = C(r_o) x(t)$.

The Laplace transform of the electric field is defined as follows

$$\tilde{E}(r_o, s) = \int_0^\infty E(r_o, t) e^{-st} dt, \quad (13)$$

from which it is easily verified that $\tilde{E}(r_o, s) = C(r_o) \tilde{x}(s)$.

Then (10) can be converted to the frequency domain by substituting the differential operator $\partial/\partial t$ with the complex frequency variable $s = \beta + j\omega$, $\beta > 0$. The electric field is then written in terms of the excitation current source as

$$\tilde{E}(r_o, s) = G(r_o, s) i(s), \quad (14)$$

with the transfer function $G(r_o, s)$ given by

$$G(r_o, s) = C^*(sI - A)^{-1} b, \quad (15)$$

where $I$ is the identity matrix. From (14) and (15) the following expression for the $x$ component of $\tilde{E}(r_o, s)$ is obtained

$$[\tilde{E}(r_o, s)]_x = [G(r_o, s)]_x i(s) \quad (16)$$

$$[G(r_o, s)]_x = L \sum_{\nu=1}^N \frac{[E_\nu(r_s)]_x [E_\nu(r_o)]_x^* s}{2W_\nu (s^2 + \omega_\nu^2)}.$$

It is observed that the transfer function in (16) has $2N$ poles all present in mirrored pairs. Furthermore, we have $2N - 1$ zeros with one zero at $s = 0$ and the remaining zeros present in mirrored pairs. These zeros can be determined from (16) by combining the sum of fractions such that we obtain a single fraction and then finding complex values of $s$ for which the numerator becomes zero. Because the cavity is lossless, the poles are always located at the imaginary frequency axis, $s = \pm j\omega_\nu$. However, the zeros can have nonzero real part, i.e. $\beta_\nu \neq 0$ in $s = \beta_\nu \pm j\omega_\nu$. Note that the locations of the zeros depend on the shape of the cavity and the location of the observation point and source, i.e. they depend on $[E_\nu(r_s)]_x$, $[E_\nu(r_o)]_x^*$, $W_\nu$ and $\omega_\nu$. As a result, the transfer function of the $x$, $y$ and $z$ component of $E(r_o, s)$ share the same poles, however their zeros may be different.

## 3. Phase reconstruction method

### 3.1. Kramers-Kronig relations for magnitude and phase

We assume the transfer function of a system has a causal time response. The causal response function in Fourier space is then explicitly given by

$$G(\omega) = \int_0^\infty g(t) e^{-j\omega t} dt, \quad (17)$$

$g(t)$ being real implies that $G(-\omega) = G^*(\omega)$.

We now construct the analytical continuation of $G$ by replacing $\omega$ with the complex frequency variable $s = \beta + j\omega$, where $\beta \geq 0$. It follows that $G(s)$ is an analytical function in the left half of the $s$-plane. The Cauchy integration theorem for the causal transfer function $G(s)$ yields [9]

$$G(s) = \frac{1}{2\pi j} \oint_C \frac{G(s')}{s' - s} ds', \quad (18)$$

where the closed contour $C$ consists of the imaginary axis and the infinite semicircle enclosing the right half of the $s$-plane. We assume that $C$ is oriented counterclockwise. Let us also assume that $G(s) \to 0$ for $|s| \to \infty$. Considering then again imaginary values of $s = \lim_{\beta \downarrow 0} \beta + j\omega$ which leads to the KK relations [9]

$$\text{Re } G(j\omega) = -\frac{1}{\pi} \mathcal{P} \int_{-\infty}^{\infty} \frac{\text{Im } G(j\omega')}{\omega' - \omega} d\omega' \quad (19)$$

$$\text{Im } G(j\omega) = \frac{1}{\pi} \mathcal{P} \int_{-\infty}^{\infty} \frac{\text{Re } G(j\omega')}{\omega' - \omega} d\omega',$$

where $\mathcal{P}$ denotes the principal value. Thus, the real and imaginary parts are Hilbert transform [10] pairs of each other

$$\text{Re } G = \mathcal{H}(\text{Im } G) \quad (20)$$
$$\text{Im } G = -\mathcal{H}(\text{Re } G).$$

Causality yields via the KK relations the imaginary part of the frequency response if the real part is known and vice versa. Therefore, one may expect that either the magnitude or the phase is enough to reconstruct analytic frequency responses. These are defined as

$$G(j\omega) = |G(j\omega)| \exp j\gamma(j\omega) \quad (21)$$

where $\gamma(j\omega)$ is the phase. For completeness, their relation to the real and imaginary part is given by

$$|G(j\omega)| = \sqrt{\left(\text{Re } G(j\omega)\right)^2 + \left(\text{Im } G(j\omega)\right)^2} \quad (22)$$
$$\tan \gamma(j\omega) = \frac{\text{Im } G(j\omega)}{\text{Re } G(j\omega)}.$$

Recalling that $G(-j\omega) = G^*(j\omega)$, an implication of $g(t) = g^*(t)$, yields the symmetry relations

$$|G(-j\omega)| = |G(j\omega)| \quad (23)$$
$$\gamma(-j\omega) = -\gamma(j\omega).$$

It is more the rule than the exception that in experiments only the magnitude can be measured. The question is whether the phase, and eventually the analytic frequency response, can be reconstructed via KK(-like) relations. The theory related to the last question is discussed in, for example, [2] where also the concept of minimum phase is introduced. Here we partly follow [3] which appears to be most useful for our purposes. The reader may consult the references for more details. In order to separate the phase, the complex logarithm of the response is studied. We introduce the infinitesimal parameter $\kappa$ and consider

$$\ln G(j\omega + \kappa) = \ln|G(j\omega + \kappa)| + j\gamma(j\omega + \kappa), \quad (24)$$

where $\kappa > 0$. At this point it seems to be possible to formulate dispersion relations for the logarithm of the magnitude and the phase. Indeed, with some caveats to be addressed below, the phase follows as

$$\gamma(j\omega) = -\mathcal{H}(\ln|G(j\omega)|) \quad (25)$$

or, explicitly showing arguments and the $\kappa$ prescription,

$$\gamma(j\omega) = \frac{1}{\pi} \mathcal{P} \int_{-\infty}^{\infty} \frac{\ln|G(j\omega' + \kappa)|}{\omega' - \omega} d\omega'. \quad (26)$$

Note that $\kappa$ ensures that the integration path does not contain any poles or zeros of $G(s)$ in case they lay on the imaginary axis. The first caveat which has to be made is that the complex response does not support zeros in the right half of the $s$-plane. In that case the logarithm is not defined and to satisfy the KK relations the appearing logarithm should be analytic for $\text{Re } s > 0$. The requirement that the logarithm of the magnitude and the phase form a Hilbert pair is called the minimum phase condition, see e.g. [2]. In [11] this minimum phase condition has been imposed. The second caveat is that the complex response does not support zeros at infinite complex frequency i.e. $|s| \to \infty, G(s) \to 0$.

### 3.2. Numerical results

To test the KK relations for magnitude and phase presented in subsection 3.1 with the cavity transfer function as defined in subsection 2.3, we present three examples here. For the different examples, we vary the location where the electric field is observed, $r_o$. Also, the dimensions of the cavity are changed. For all examples the magnitude of the transfer function is used to reconstruct the phase using (26). We compare the actual phase of the transfer function to the reconstructed phase and see if the minimum phase condition is violated, which is dependent on the location of the zeros in the transfer function. Note that there is also a zero at infinite complex frequency because $[\boldsymbol{G}(\boldsymbol{r}_o, s)]_x \to 0$ for $|s| \to \infty$, as is observed from (16). For all examples studied here however, it has been observed that the phase error caused by the zero at infinity is negligible. In the examples we only consider the $x$ component of the electric field, which means that we only must compute $[\boldsymbol{G}(\boldsymbol{r}_o, s)]_x$. The resulting transfer functions were verified using numerical results obtained by using the Method of Moments (MoM) solver in FEKO [12].

To observe the effect of phase errors on the time response of the electric field when using the reconstructed phase, we excite the cavity fields using a pulsed current generated by a point dipole source. The time response of the current pulse is defined as

$$i_p(t) = E_0 k_1 \exp(-a_1[t - t_d]) \quad (27)$$
$$- \exp(-b_1[t - t_d]),$$

with $E_0 k = 6.5 \times 10^4$ V/m, $a_1 = 4 \times 10^7$ s$^{-1}$, $b_1 = 6 \times 10^8$ s$^{-1}$ and $t_d = 1 \times 10^{-8}$ s. The current pulse, $i_p(t)$, is subsequently filtered through a first order high pass filter with crossover frequency of 0.6 MHz, such that the resulting current source inside the cavity, $i(t)$, has zero mean. The pulse has a -3 dB frequency bandwidth of approximately 8 MHz and a -10 dB frequency bandwidth of approximately 20 MHz.

We then compare the time response of the electric field using the directly calculated phase to the time response of the electric field using the reconstructed phase. For the first example we set the cavity size parameters in *Fig. 1* to $a = 0.8$ m, $b = 0.9$ m and $d = 1$ m. The source is positioned at $\boldsymbol{r}_s = [0 \quad b/3 \quad d/3]$ and the observation point is located at $\boldsymbol{r}_o = [a/3 \quad b/3 \quad d/3]$. For all the following calculations, $\kappa$ in (26) is set to $0.25/\pi$ MHz . This value was experimentally determined to give good results. Note that the length of the dipole source as defined in (5) only scales the transfer function by a constant factor while the phase response is not effected. For calculating $[\boldsymbol{G}(\boldsymbol{r}_o,s)]_x$, all modes for which $\omega_v/2\pi > 500$ MHz are discarded, which means that a total of $N = 32$ (16 TE and 16 TM) modes are used for the calculations.

We first compare the phase directly extracted from the model with the reconstructed phase. In *Fig. 2-a* we see that the reconstructed phase starts to deviate for frequencies higher than 430 MHz, which is due to the fact that two zeros are present in the right half of the *s*-plane at $s = 2\pi \times (0.055 \pm j4.4) \times 10^8$ . These zeros violate the minimum phase condition needed for (26) to hold.

The resulting time response is shown in *Fig. 2-b*. We see that the reconstructed time response is very similar to the directly calculated time response. This is because the bandwidth of the pulse is relatively small, thus the cavity modes for which the reconstructed phase is starting to deviate have a much smaller magnitude than the lower frequency modes which are excited more significantly by the pulse.

As a second example, we take the observation point at $\boldsymbol{r}_o = 2[a/3 \quad b/3 \quad d/3]$. The reconstructed phase is shown in *Fig. 3-a*. Similar effects are observed compared to the previous case shown in *Fig. 2*. However, for this case a total of six zeros are present in the right half of the *s*-plane, which are now located at: $s = 2\pi \times (2.3 \pm j3.5) \times 10^8$, $s = 2\pi \times (0.061 \pm j4.3) \times 10^8$  and  $s = 2\pi \times (0.31 \pm j5) \times 10^8$ . The zeros of the transfer function are located at different points compared to the previous case, since we changed the observation point, $\boldsymbol{r}_o$, which also moves some of the zeros further away from the $j\omega$ axis in the right half of the *s*-plane. This, combined with the fact that there are significantly more zeros in the right half of the *s*-plane, explains why the reconstructed phase is deviating much more severely compared to the previous case.

The resulting time response shows a much larger difference between the reconstructed and directly calculated results, as is presented in *Fig. 3-b*. This is as expected, because the difference between the reconstructed phase and the directly calculated phase is already in the order of 90° at the lowest order mode resonance frequency, i.e. 220 MHz. However, as the lowest order mode has the largest contribution to the resulting time responses, a simple phase correction of 90° should already significantly improve the reconstructed time response. We see that apart from this 90° phase shift, the overall shape and most importantly the envelope and peak values of the time signal are preserved.

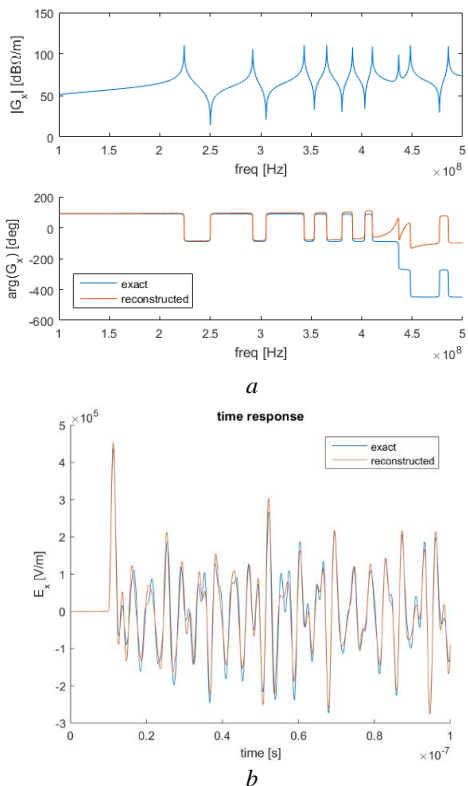

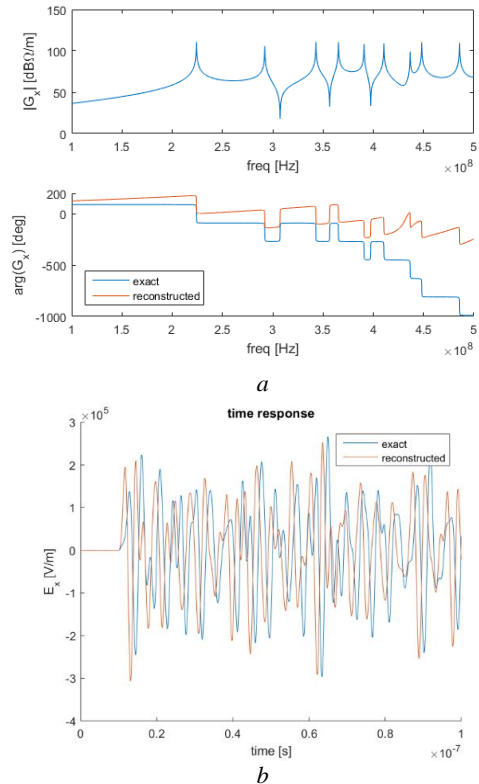

***Fig. 2.*** *(a) Magnitude, phase and reconstructed phase of $G_x(j\omega)$ as a function of frequency for $\boldsymbol{r}_o = [a/3 \quad b/3 \quad d/3]$. (b) Time response of $E_x$.*

***Fig. 3.*** *(a) Magnitude, phase and reconstructed phase of $G_x(j\omega)$ as a function of frequency for $\boldsymbol{r}_o = 2[a/3 \quad b/3 \quad d/3]$. (b) Time response of $E_x$.*

From this we conclude that some error in the reconstructed phase might be allowed. However, this error largely depends on the locations of the zeros of the transfer function. In practice this means that we do not know what phase correction we must apply, unless we are able to directly determine the locations of the complex zeros.

As third example, we increase the size of the cavity and change its shape by setting $a = 8$ m, $b = 9$ m and $d = 3$ m. The location of the source and the observation point remain unchanged from the second example. Note that because the cavity is much larger now, the modes have much larger wavelengths and as a result, the corresponding resonance frequencies are much lower. Thus, more modes are excited by the EMP and we need to consider a larger number of modes in the mode expansion. To bound computation time, we limit the number of modes used in the mode expansion to 50 TE- and 50 TM-modes, hence resulting in a total of $N = 100$ modes. The results for the configuration are shown in *Fig. 4*. Note that the transfer function is now plotted between 40 MHz and 130 MHz.

In *Fig. 4-a* we see that the discrepancy between the exactly calculated and reconstructed phase is now more severe compared to the cases shown in *Fig. 2-a* and *Fig. 3-a*. This is caused by the fact that now a total number of 16 zeros are located in the right half of the *s*-plane compared to only 2 zeros for the example shown to *Fig. 2* and 6 zeros for the example shown in *Fig. 3*.

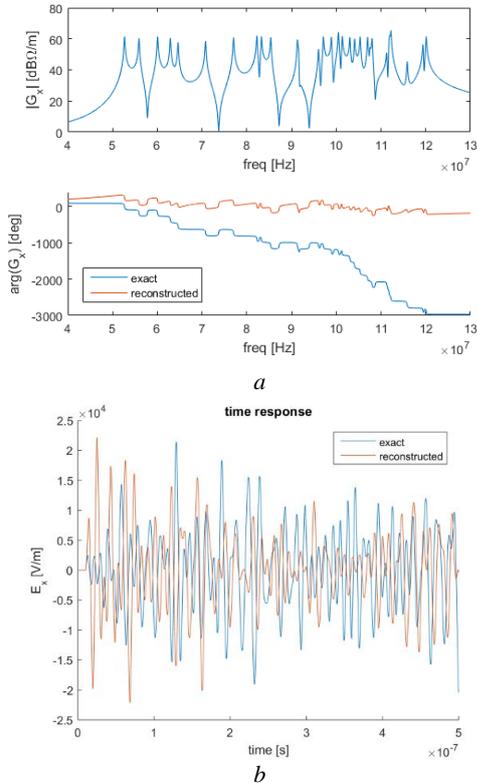

***Fig. 4.*** *(a) Magnitude, phase and reconstructed phase of $G_x(j\omega)$ as a function of frequency for $\mathbf{r}_o = 2[a/3 \quad b/3 \quad d/3]$. (b) Time response of $E_x$. A larger cavity was used here.*

Next, we calculate the resulting time response of the electric field $E_x(t, \mathbf{r}_o)$. The result is shown in *Fig. 4-b* where we compare the time responses calculated using the exact phase and the reconstructed phase. We clearly observe the discrepancy between the exact and reconstructed phase and exact and reconstructed time response, which is caused by the zeros lying in the right half of the *s*-plane and the zero and infinite complex frequency. For this case, a simple constant phase shift is not sufficient, and we need an advanced method to correct the phase error.

## 4. Phase correction using the Blaschke product

### 4.1. Blaschke product and modified Kramers-Kronig relations

At this point, we resume the theory on KK relations for magnitude and phase [3]. We address the possibility of zeros of the response function in the right half of the *s*-plane. The logarithm of the magnitude is not defined at such zeros. The symmetry relations, after analytic continuation for complex $s = \beta + j\omega$, are written as

$$G(j\omega + \beta) = G^*(j\omega - \beta). \tag{28}$$

It implies that zeros of $G$ appear in pairs in the *s*-plane. Zeros with a vanishing real part form an exception. There may also be zeros for infinite complex frequency. In order to deal with the zeros of the response at finite frequency, we follow [3] which has been preceded by the seminal papers [13] [14]. It is specifically assumed that the complex numbers $z_n, n = 1, \dots N$, with $\text{Re } z_n > 0$ are zeros of $G$. The Blaschke product is then defined as

$$B(s) = \prod_{n=1}^{N} \frac{s - z_n}{-z_n^* - s}. \tag{29}$$

Next the modified response function, $\hat{G}(s)$, is constructed as

$$G(s) = \hat{G}(s)B(s). \tag{30}$$

The expansion of $\hat{G}(s)$ is possible if the integral satisfies

$$\int_0^\infty \frac{\ln|\hat{G}(j\omega' + \kappa)|}{1 + \omega^2} d\omega' < \infty, \tag{31}$$

i.e. converges. The Blaschke product should converge as well, which can be reformulated as [13]

$$\sum_{n=1}^{N} \frac{\text{Re } z_n}{|z_n|^2} < \infty. \tag{32}$$

This condition is obviously fulfilled for finite $N$. The magnitude of the responses $G(s)$ and $\hat{G}(s)$ are equal

$$|G(j\omega + \kappa)| = |\hat{G}(j\omega + \kappa)|. \tag{33}$$

It can be verified by first recalling that the zeros of $\hat{G}(s)$ in the right half $s$-plane appear in complex conjugated pairs. We can therefore write the product for real $\omega$ as

$$B(j\omega) = \prod_{k=1}^{N/2} \frac{j\omega - z_k}{j\omega + z_k^*} \frac{j\omega - z_k^*}{j\omega + z_k}, \quad (34)$$

with index $k$ labelling the pairs. Since

$$\left|\frac{j\omega - z_k}{j\omega + z_k^*} \frac{j\omega - z_k^*}{j\omega + z_k}\right|^2 \quad (35)$$
$$= \left(\frac{j\omega - z_k}{j\omega + z_k^*} \frac{j\omega - z_k^*}{j\omega + z_k}\right)\left(\frac{j\omega - z_k}{j\omega + z_k^*} \frac{j\omega - z_k^*}{j\omega + z_k}\right)^* = 1,$$

we get $|B(j\omega)| = 1$. Hence, the validity of relation (33) is determined.

The modified response function $\hat{G}(s)$ by applying the Blaschke product has by construction no zeros in the right half of the $s$-plane. Therefore, we obtain the KK relations for its magnitude and phase $\hat{\gamma}$

$$\hat{\gamma}(j\omega) = \frac{1}{\pi}\mathcal{P}\int_{-\infty}^{\infty} \frac{\ln|\hat{G}(j\omega' + \kappa)|}{\omega' - \omega} d\omega' \quad (36)$$
$$= \frac{1}{\pi}\mathcal{P}\int_{-\infty}^{\infty} \frac{\ln|G(j\omega' + \kappa)|}{\omega' - \omega} d\omega'.$$

The phases of the responses are related by

$$\gamma(j\omega) = \hat{\gamma}(j\omega) + \arg B(j\omega), \quad (\mathrm{mod}\ 2\pi). \quad (37)$$

where $\arg B(j\omega)$ can be seen as a phase correction term which corrects the phase error caused by zeros of $G(j\omega)$ located in the right half of the $s$-plane. The $\arg B(j\omega)$ term is obtained by first calculating

$$\arg\left(\frac{z_k - j\omega}{z_k^* + j\omega}\right) = \quad (38)$$
$$\arctan2(2x_k(y_k - \omega), x_k^2 - (y_k - \omega)^2),$$

where $z_k = x_k + jy_k$; $x_k, y_k \in \mathbb{R}$; $x_k > 0$. Secondly, we get

$$\arg\left(\frac{z_k^* - j\omega}{z_k + j\omega}\right) = \quad (39)$$
$$-\arctan2(2x_k(y_k + \omega), x_k^2 - (y_k + \omega)^2),$$

which eventually yields

$$\arg B(j\omega) = \quad (40)$$
$$\sum_{k=1}^{N/2}[\arctan2(2x_k(y_k - \omega), x_k^2 - (y_k - \omega)^2)$$
$$- \arctan2(2x_k(y_k + \omega), x_k^2 - (y_k + \omega)^2)].$$

The complete modified KK relations are written as

$$\gamma(j\omega) = \frac{1}{\pi}\mathcal{P}\int_{-\infty}^{\infty} \frac{\ln|G(j\omega' + \kappa)|}{\omega' - \omega} d\omega' + \arg B(j\omega) \quad (41)$$

and

$$\ln|\hat{G}(j\omega + \kappa)| \quad (42)$$
$$= \frac{1}{\pi}\mathcal{P}\int_{-\infty}^{\infty} \frac{\arg B(j\omega') - \gamma(j\omega')}{\omega' - \omega} d\omega'.$$

In practice, the locations of the $N$ zeros may actually not be known. In Section 5 we propose a method which enables us to find the locations of the zeros using only amplitude measurements of the time response while using specific excitation signals. This allows us to calculate the phase correction term $\arg B(j\omega)$ in (37), needed to get the correctly reconstructed phase for the transfer function.

### 4.2. Additional modifications for zeros at infinity

In subsection 3.1 we discussed that transfer functions which have a zero at infinite complex frequency i.e. $|s| \to \infty$, $G(s) \to 0$, do not satisfy the KK relations for magnitude and phase. Here we propose to introduce once more a Blaschke factor for zeros at infinity. We first assume that possible other zeros are already being taken care of by the procedure given in subsection 4.1. Next, we assume that

$$G(s) \sim \frac{1}{s^k} \quad \text{for} \quad |s| \to \infty, \quad (43)$$

for some positive integer $k$. Analogous to the use of the Blaschke product, we define a modified response

$$\bar{G}(s) = G(s)P(s). \quad (44)$$

These relations yield for the complex logarithm

$$\log \bar{G}(s) = \log|G(s)| + \log|P(s)| + j\arg G(s) \quad (45)$$
$$+ j\arg P(s) \quad (\mathrm{mod}\ 2\pi).$$

At this point we specify the Blaschke factor as

$$P(s) = (s+1)^k, \quad (46)$$

which has a multiple zero at $s = -1$ but none in the right half of the $s$-plane. Furthermore, it is easily seen that

$$P(s) \sim |s|^k \quad \text{for} \quad |s| \to \infty \quad (47)$$

and consequently

$$\bar{G}(s) \sim 1 \quad \text{for} \quad |s| \to \infty. \quad (48)$$

Therefore, we obtain the KK relation for the modified response $\bar{G}$

$$\arg \bar{G}(j\omega) = \frac{1}{\pi}\mathcal{P}\int_{-\infty}^{\infty} \frac{\log|\bar{G}(j\omega' + \kappa)|}{\omega' - \omega} d\omega'. \quad (49)$$

For imaginary arguments of $P(s)$ we get

$$\log|P(j\omega)| = \frac{1}{2}k\log(\omega^2 + 1) \quad (50)$$

and

$$\arg P(j\omega) = k \arctan2(\omega, 1) \pmod{2\pi}. \tag{51}$$

Hence, we obtain the modified KK relation

$$\arg G(j\omega) = \tag{52}$$
$$\frac{1}{\pi}\mathcal{P}\int_{-\infty}^{\infty}\frac{\ln|G(j\omega' + \kappa)| + \frac{1}{2}k\log(\omega'^2 + 1)}{\omega' - \omega}d\omega'$$
$$- k \arctan2(\omega, 1) \pmod{2\pi}.$$

When also incorporating the Blaschke product and corresponding phase correction for zeros laying in the right half of the *s*-plane, the modified KK relation finally becomes

$$\gamma(j\omega) = \arg G(j\omega) + \arg B(j\omega). \tag{53}$$

where $\arg B(j\omega)$ is defined in (40).

## 5. Method for obtaining zero locations using time domain measurements

In subsection 2.3 it is noted that the cavity transfer function in some cases does not satisfy the minimum phase condition. In that case a severe phase error can emerge when the unmodified KK relations are used to reconstruct the phase response from the magnitude of the frequency response. The locations of the complex zeros within the right half of the *s*-plane need to be determined in order to correct errors in the reconstructed phase by using modified KK relations, as discussed in subsection 4.1. In this section we propose a method to estimate the locations of the complex zeros in the right half of the *s*-plane using specific excitation signals and measuring the resulting time responses of the system.

We define a transfer function of a system in the *s* domain

$$G(s) = \frac{Y(s)}{X(s)}. \tag{54}$$

The system has input $X(s)$ and output $Y(s)$, and represents an arbitrary linear time-invariant single-input single-output system, such as a cavity with input electric field strength of an excitation source and as output the measured electric field strength. Let us further assume that our zero-search input signal in the time domain $x(t)$ is given by

$$x(t) = u(t)u(-t + t_{off})f(t) \tag{55}$$
$$f(t) = (\exp(\beta t) - 1)\cos(\omega t),$$

where $u(t)$ is the Heaviside step function. With $\beta > 0$, $f(t)$ is an exponentially *increasing* cosine wave, which is made causal by multiplication with $u(t)$. The zero-search input signal $x(t)$ is switched off at $t = t_{off}$ by multiplication with $u(-t + t_{off})$, which ensures that the Laplace transform of $x(t)$ exists for $\beta > 0$. The function $f(t)$ is defined such that the signal is zero for $t = 0$.

The one-sided Laplace transform of $f(t)$ is given by

$$F(s) = \frac{s - \beta}{(s - \beta)^2 + \omega^2} - \frac{s}{s^2 + \omega^2}, \tag{56}$$

which has poles at $s = \beta \pm j\omega$ and $s = \pm j\omega$. Note that the Laplace transforms of $f(t)$ and $x(t)$ have equal poles. When the first pole pair is placed at $s = \beta_k \pm j\omega_k$, with $\omega_k$ and $\beta_k$ corresponding to the *k*-th zero pair of the transfer function, the *k*-th zero pair of the transfer function and the corresponding pole pair of the zero-search input signal cancel each other. In the time domain this can be detected by analysing the output of the system, $y(t; \beta, \omega)$, and observing if the $\exp(\beta t)\cos(\omega t)$ term in $x(t)$ is suppressed by the zero present in the transfer function.

We will illustrate this using a simple example. Let us define the transfer function

$$G(s) = \frac{s - 1}{s^2 + 1}, \tag{57}$$

which has one real zero at $s = 1$, and one pole pair at $s = \pm j$. Note that the zero is positioned in the right half of the *s*-plane on the real frequency axis, and as a result this is a non-minimum phase system. As a zero-search input signal we use $x(t)$ from (55) with $\omega = 0$ in order to find the location of the zero on the real frequency axis. After performing the inverse Laplace transform on $G(s)X(s)$ we obtain the system response $y(t; \beta, \omega = 0)$ to the input $x(t)$ for $0 < t < t_{off}$

$$y(t; \beta, \omega = 0) = \frac{(\beta^2 - 1)\exp(\beta t)}{(\beta^2 + 1)} - \tag{58}$$
$$\frac{\beta^2(\cos(t) + \sin(t)) + \beta(\cos(t) - \sin(t))}{(\beta^2 + 1)}.$$

Note that for $\beta = 1$ the first term in $y(t; \beta, \omega = 0)$ containing the zero-search input signal $\exp(\beta t)$, becomes zero. This is because $F(s)$ in (56) has a pole at $s = \beta$ which, for $\beta = 1$, cancels the zero at $s = 1$ in $G(s)$. Thus, we sweep $\beta$ until we observe the first term, containing $\exp(\beta t)$, to become zero in $y(t; \beta, \omega = 0)$ and thereby the location of the zero is determined. The fact that $\exp(\beta t)$ increases exponentially in time makes this relatively easy to detect in practice, e.g. by measuring the magnitude of the system response. The $\exp(\beta t)\cos(\omega t)$ signal is detected in the output by integrating the magnitude of the system response over time as follows

$$E_{int}(\beta, \omega) = \int_{t=0}^{t_{off}} |y(t; \beta, \omega)| dt. \tag{59}$$

When the exponential function present in the output $y(t; \beta, \omega)$ is suppressed by the (complex) zero in the transfer function, we observe a decrease in $E_{int}(\beta, \omega)$ for a certain combination of $\beta$ and $\omega$. This minimum provides us the estimated location of the pertaining complex zero.

Due to the exponential term in (55), the input $x(t)$ will increase rapidly when $\beta$ increases. To ensure that the output signal $y(t; \beta, \omega)$ does not becomes too large numerically we scale the input according to

$$\hat{x}(t) = \exp(-\gamma\beta)\, x(t), \qquad (60)$$

where $\gamma$ is a positive constant.

### 5.1. Blind tests for verification of the zero-search method

We now test the zero-search method on transfer functions having one or more zero pairs in the right half of the *s*-plane at varying locations. To enforce our confidence in the method, we perform several blind tests. This means that no a priori knowledge about the zero locations is used for finding them. The blind testing method used here is done through the following procedure: The second author has a transfer function of which only he knows the locations of the zeros in the right half of the *s*-plane. The second author then calculates the impulse response of the transfer functions and hands it to the first author. Note that the first author only has the numerically calculated impulse response of the system and no information about the corresponding transfer function, thus he does not know the locations of the zeros. Note that for computing $y(t; \beta, \omega)$ in (59) the convolution between the impulse response and the zero-search input signal defined in (55) can be used. The first author then uses the zero-search method described in Section 5 to attempt finding the locations of the zeros in the right half of the complex *s*-plane.

The transfer function used in these blind tests is defined as

$$G(s) = \frac{i}{s+\chi} \prod_k \frac{s+\mu_k}{s-\mu_k} \frac{s+\mu_k^*}{s-\mu_k^*}, \qquad (61)$$

where $k$ labels the number of pairs and $\mu_k = \beta_k + j\omega_k$ with real and positive $\beta_k, \omega_k, \chi$. This form of the transfer function is chosen such that all zeros are in the right half of the *s*-plane are mirrored across the $j\omega$ axis with respect to the locations of the poles. It is very similar to the product expansion of the causal *S*-matrix describing scattering of the Maxwell field [13].

We discuss the result of one of the blind tests in detail. The example contains two zero pairs in the right half of the *s*-plane where the parameter values used are $\beta_1 = 0.65$, $\omega_1 = 5$, $\beta_2 = 1.3$, $\omega_2 = 10$ and $\chi = 1$. This results in the zeros of (61) being located at $s = 0.65 \pm 5j$ and $s = 1.3 \pm 10j$. For the analysis we set $t_{off} = 10[s]$. The inverse exponential scaling constant is set to $\gamma = 10$, which ensures that the input signal $|x(t_{off})|$ never exceeds 1. In *Fig. 5*, $E_{int}$ as defined in (59) is plotted for varying $\beta$ and $\omega$. The estimated location of the first zero is $(\beta = 0.652, \omega = 5.01)$, which is close to the actual location. Next, we observe the location of the second zero estimated at $(\beta = 1.29, \omega = 9.98)$ which is also close to the actual location. As the zeros form complex conjugated pairs, we only need to localize 2 out of 4 zeros. The accuracy of the estimated zero locations here is mainly limited by the sample density of $\beta$ and $\omega$ used to find the zeros.

More of these cases, with either one or two zero pairs were blindly tested for varying values of $\beta_k, \omega_k, \chi$. All these tested cases provided similar accuracy in terms of the estimated zero locations. Thus, we can conclude that this method looks very promising for finding the zero locations without any a priori knowledge of the transfer function.

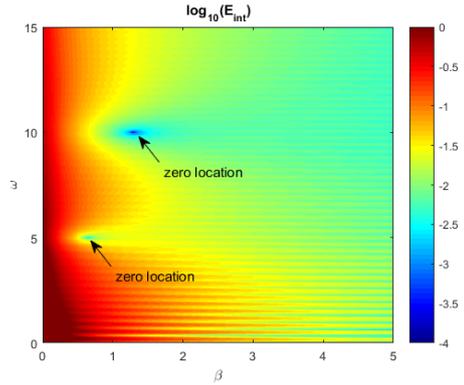

*Fig. 5.* $E_{int}$ plotted as a function of $\beta$ and $\omega$.

### 5.2. Numerical results for the cavity transfer function

As a next step, we test our method to search for the complex zeros of the cavity transfer function defined in (16). The zero-search method is first applied to the example shown in *Fig. 3*. The magnitude of the transfer function for the *x* component of the electric field is displayed in *Fig. 6-a*. $E_{int}$ is plotted as a function of $\beta$ and $\omega$ in *Fig. 6-b*

After visual inspection, three complex zeros located in the right half of the *s*-plane are identified. Other zeros are located on the $j\omega$ axis, as can be seen e.g. around $(\beta = 0, \omega = 2 \times 10^9)$, but these do not pose a problem as we use the parameter $\kappa$ to ensure that these zeros are positioned left of the integration path in (26). Note that all zeros come in complex pairs, mirrored around the $\beta$ axis.

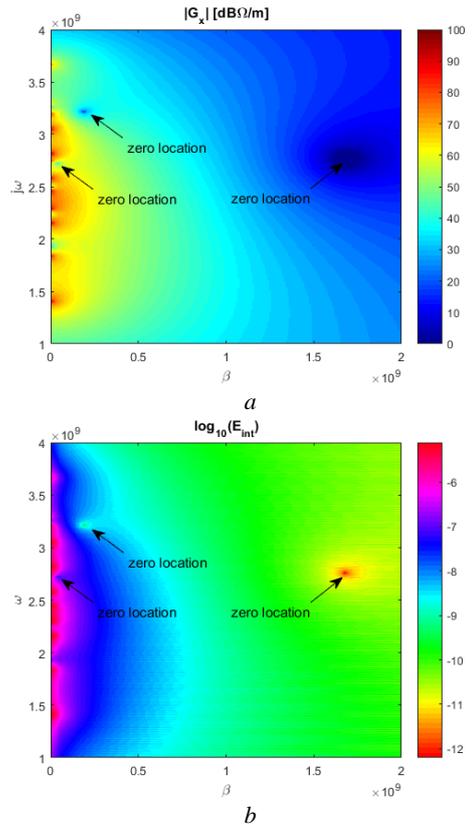

*Fig. 6.* (a) The magnitude response of $G_x(s)$ in db$\Omega/m$ in the complex *s*-plane. (b) $E_{int}$ plotted as a function of $\beta$ and $\omega$.

We aim to find the locations of the zeros located in the right half of the $s$-plane, which are marked by an arrow in *Fig. 6-a*. To find the locations of these zeros we excite the cavity using the input function $x(t)$ defined in (55). The output signal, $y(t; \beta, \omega)$, is then analysed by calculating the integral for $E_{int}(\beta, \omega)$ as defined in (59).

To this end, we will sweep both $\beta$ and $\omega$. The time duration of the excitation is set to $t_{off} = 200$ ns. The inverse exponential scaling constant is set to $\gamma = 200 \times 10^{-9}$, which ensures that $|x(t_{off})|$ never exceeds 1.

The results are shown in *Fig. 6-b*. We clearly observe that the locations of the zeros are marked by a decrease in $E_{int}(\beta, \omega)$. Comparing *Fig. 6-a* to *Fig. 6-b* we see that the decrease in $E_{int}(\beta, \omega)$ appears at exactly the same place where the transfer function becomes zero in the $s$-plane. Hence, for this case the zeros can be accurately located by varying $\beta$ and $\omega$ and observe the local minima of $E_{int}(\beta, \omega)$ where the gradient of $E_{int}(\beta, \omega)$ becomes zero. Note that because the zeros form complex conjugated pairs, the other three zeros for negative $j\omega$ are now also automatically found.

All three zero pairs, present in the right half of the $s$-plane, are determined. The accuracy of our localisation method is mainly limited by the sample density of $\beta$ and $\omega$ used to find the zeros. With the zero locations known, the corrected phase is calculated using (53) and the result in *Fig. 7* is obtained.

Comparing *Fig. 7* with *Fig. 3-a* we observe that the Blaschke product does indeed correct the phase error caused by zeros located in the right half of the $s$-plane and the phase error caused by the zero at infinite complex frequency. Thus, we demonstrated that by using modified KK relations, the phase can be successfully determined from the magnitude of the frequency response.

It is verified that the phase error caused by the zero at infinite complex frequency is negligible within the bandwidth considered here, and thus (53) and (41) provide similar numerical results. This is likely due to the large number of higher order modes, which causes the magnitude of the complex transfer function to go to zero slowly for increasing frequencies. Other cases however might benefit from using (53) instead of (41) when the magnitude of the complex transfer function goes to zero more rapidly for increasing frequencies.

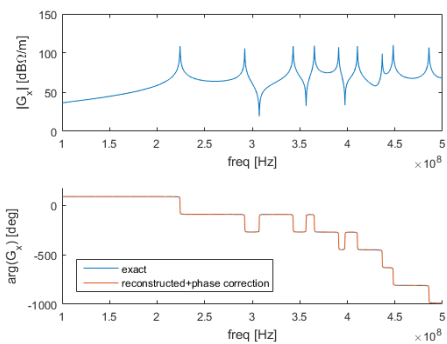

*Fig. 7. Magnitude of $G_x(j\omega)$ and the reconstructed phase of $G_x(j\omega)$ corrected using the Blaschke product (modified KK relations).*

## 6. Noise sensitivity

In this section we investigate the influences of noise on the accuracy of the proposed zero-search method and the phase reconstruction method using the modified KK relations. This will be illustrated using the cavity transfer function defined in (16) for the example shown in *Fig. 4*.

To apply the proposed phase reconstruction method using the modified KK relations in combination with the zero-search method, two separate measurements must be performed. The first measurement is performed in the time domain, where the time responses of the system are measured to find the zeros in the right half of the $s$-plane. The second measurement is performed in the frequency domain where the magnitude of the frequency response of the system is measured, which is used to reconstruct the phase using the modified KK relations. In practice, both measurements will be subject to measurement noise.

First, we will simulate the effects of measurement noise on the zero-search method. Gaussian noise with zero mean and a standard deviation of $10^{-4}$ V/m is added to the measured time response, $y(t; \beta, \omega)$, in (5). The magnitude of the transfer function for the $x$ component of the electric field is displayed in *Fig. 8-a*. $E_{int}$ is plotted as a function of $\beta$ and $\omega$ in *Fig. 8-b*.

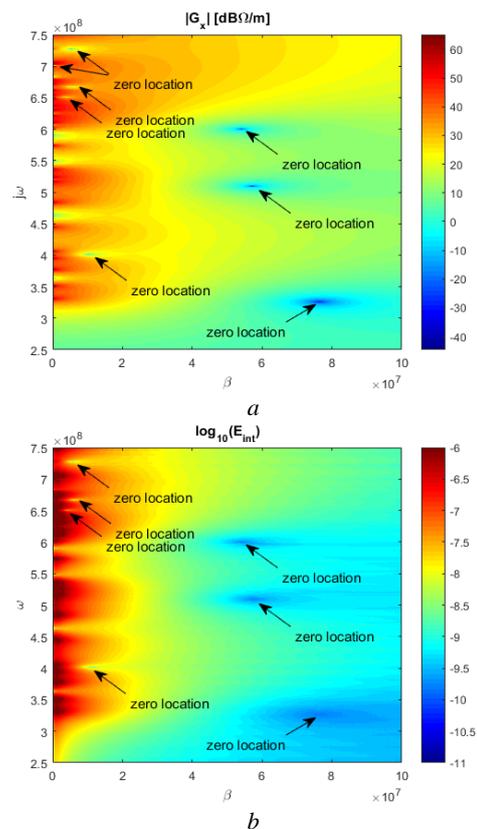

*Fig. 8. (a) The magnitude response of $G_x(s)$ in db$\Omega/m$ in the complex $s$-plane. (b) $E_{int}$ plotted as a function of $\beta$ and $\omega$.*

We observe that all zeros except the zero at ($\beta = 0.1 \times 10^6, \omega = 7 \times 10^8$) can be located with high accuracy. The zero that cannot be located using our zero-search method is positioned very close to the poles on the imaginary frequency axis. This causes the influence of this zero to be nearly cancelled by a nearby pole, making locating the zero using the measured time response very difficult. Note that we consider a lossless cavity here. In practice the poles will usually be positioned to the left of the imaginary frequency axis due to electromagnetic losses in the cavity. Our hypothesis is that this should make locating zeros close to the imaginary frequency axis easier in practice than encountered in this exercise.

In the example shown here, the noise only influences the accuracy of locating the zeros positioned far away from the imaginary frequency axis, as the measured signal strength is very low around these zeros. The average signal to noise ratio over time (RMS) of the measured time response is approximately 2 when $\beta = 7.7 \times 10^7$ and $\omega = 3.3 \times 10^8$, which is just sufficient to be able to locate the corresponding zero. When the noise floor is too high, zeros cannot be located as the they disappear below the noise floor. This problem can be circumvented by applying a stronger excitation signal to the system, e.g. by decreasing $\gamma$ in (60).

Next, we simulate the effects of measurement noise on the phase reconstruction method using the modified KK relations. Complex Gaussian noise with zero mean and a standard deviation of 3.16 $\Omega$/m for both real and imaginary part is added to the frequency response, $G(j\omega' + \kappa)$, in (41). With all but one of the zero locations known, the corrected phase is calculated using (53) and the result in *Fig. 9* is obtained. Note that all noise outside the frequency band of interest, 50 MHz---125 MHz, is filtered out using a bandpass filter.

Comparing the result in *Fig. 9* to the result in *Fig. 4-a* we see that, despite the fact that noise has been added here, the phase error is significantly reduced by using modified KK relations. We see 5 distinct locations where the absolute phase error becomes larger than 20°. The first 4 spikes in the absolute error are caused by a low signal to noise ratio. However, at these locations the transfer function is very small and thus these errors will not have a significant impact on the reconstructed time response.

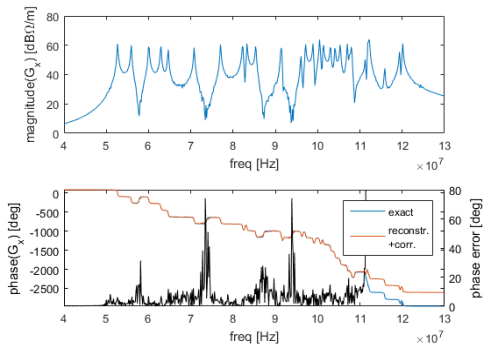

*Fig. 9.* Magnitude of $G_x(j\omega)$ and the reconstructed phase of $G_x(j\omega)$ corrected using the Blaschke product (modified KK relations). Also, the absolute error of the reconstructed phase is displayed.

The last error spike around 110 MHz is caused by one of the missing zero's in the correction factor, which we could not find using the proposed zero-search method. The effect on the reconstructed time response depends on the bandwidth of the input signal as only frequencies above 110 MHz are affected.

In this section we aimed to apply a realistic amount of measurement noise to determine the noise sensitivity of our zero-search and phase reconstruction method. In future research we need to further investigate the influences of noise on our methods and draw up requirements for measuring equipment used for performing experiments.

The localisation of unknown zeros of a transfer function using time domain responses, and the subsequent reconstruction of the phase response may be interpreted as an inverse problem. The locations of the zeros as well as the number of zeros in the right half of the *s*-plane are, in our approach, parameters which need to be estimated appropriately. Note that we search for local minima in $E_{int}$, indicating the zero locations, and not for a global minimum regarding an inverse problem. In practice however, false minima might occur, e.g. due to measurement noise or errors. As a result, the system might seem to have zeros that do not exist. More research is envisioned on this topic. In the examples discussed here we did not observe this problem. To solve an inverse problem, normally one would define a cost function or an associated error norm which needs to be minimized. Here, we did not yet define an error norm to localize the zeros but use a visual approach instead.

## 7. Conclusions and recommendations

The electromagnetic field inside a PEC cavity is efficiently modelled using an eigenmode expansion when the EM fields can be sufficiently well approximated using a reasonable number of modes. With this eigenmode expansion we determine a transfer function from an excitation point dipole source to an observation point of the electric field. The locations of the zeros of this transfer function, within the complex *s*-plane, depend on the shape of the cavity and the locations of the observation point and excitation source.

If only the magnitude of the transfer function is known (by measurements), the KK relations may be used to reconstruct the phase from the magnitude. This method provides correct results when the minimum phase condition is satisfied, which is equivalent to the requirement that the transfer function has no zeros in the right half of the *s*-plane. We have observed that the transfer function of a cavity, as described here, may exhibit several zeros in the right half of the *s*-plane. As a result, the minimum phase condition is violated. This introduces a substantial error in the reconstructed phase when using the KK relations, which in turn introduces an error in the reconstructed time response of the electric field. However, we have observed that for some cases the overall shape and most importantly the envelope and peak values of the reconstructed time signal are preserved compared to the exact analytical calculations, despite the minimum phase condition being violated.

In other cases, the phase error present in the reconstructed phase is not acceptable. In order to handle these cases we used the Blaschke product, from which modified KK relations are derived. The Blaschke product allows for calculating a phase correction term which corrects the errors introduced due to the violation of the minimum phase condition. This modification requires the locations of the zeros in the right half of the *s*-plane to be accurately known. If the locations of these zeros are known, the Blaschke product and corresponding phase correction can be calculated. Subsequently, the reconstruction of the phase using modified KK relations is performed appropriately. In our numerical studies, however, within the limited bandwidth considered here, the influence of zeros at infinity turned out to be negligible in the case of the cavity transfer function.

Generally, the locations of the zeros of the transfer function in the right half of the *s*-plane are unknown, if only the magnitude of the frequency response is measured. However, it is possible to locate these zeros by analysing a specific time response of the system. When we choose the zero-search input signal such that it has a pole that coincides with a zero of the transfer function, a (significant) decrease in energy in the measured time response can be observed. This is due to the fact that a pole of the zero-search input signal gets cancelled by a zero of the transfer function. To this end an exponentially increasing cosine zero-search input signal is used as defined in (55). By varying the exponential growth constant $\beta$ and the oscillation frequency $\omega$ of the zero-search input signal, the zero locations in the right half *s*-plane of the transfer function are obtained. The usability of this approach was tested by using a blind testing method where only the time response of the system is given to the first author. These blind tests provided positive results for all verified cases, as all zeros were located successfully without a priori knowledge about the transfer function.

Additionally, the method is applied to the cavity transfer function in two examples. For the first example, all of the zeros in the right half of the *s*-plane are located, and the phase is accurately calculated from the magnitude of the frequency response by using modified KK relations. For the second example, one of the zeros could not be located because it is positioned too close to poles on the imaginary frequency axis. Our hypothesis is that in real world measurements, locating zeros close to the imaginary frequency axis is easier since losses are present and therefore poles are positioned to the left of the imaginary frequency axis.

Measurement noise will have an impact on the accuracy of the zero-search and phase reconstruction method. However, under realistic noise conditions the methods still give good results, as was demonstrated for the cavity transfer function example here.

It is important to indicate that this paper demonstrates a proof of concept for applying modified KK relations to non-minimum phase systems. More research is currently being invested by the authors in the practical realisation of the method and measurements have been planned for this purpose. This will also incorporate more in-depth research on the influences of measurement noise.

## 8. Acknowledgements

The authors would like to thank Vincent Jonker BSc for helpful discussions and support. This work is part of a research program of the Electronic Defence department of the "Nederlandse Organisatie voor Toegepast Natuurwetenschappelijk Onderzoek" (TNO) and is financially supported by the "Defensie Materieel Organisatie" (DMO) of the Dutch Ministry of Defence.